\documentclass[preprint,showpacs,preprintnumbers,nofootinbib,amsmath,amssymb]{revtex4-1}
\usepackage{graphics,epsfig,subfigure}
\usepackage{epstopdf}
\usepackage{color}

\begin{document}
\renewcommand{\baselinestretch}{1.3}
\newcommand\be{\begin{equation}}
\newcommand\ee{\end{equation}}
\newcommand\ba{\begin{eqnarray}}
\newcommand\ea{\end{eqnarray}}
\newcommand\nn{\nonumber}
\newcommand\fc{\frac}
\newcommand\lt{\left}
\newcommand\rt{\right}
\newcommand\pt{\partial}
\newcommand\tc{\textcolor[rgb]{1,0,0}}

\title{Probing extra dimension through gravitational wave observations of compact binaries and their electromagnetic counterparts}

\author{Hao Yu$^1$\footnote{yuh13@lzu.edu.cn},
        Bao-Min Gu$^1$\footnote{gubm15@lzu.edu.cn},
        Fa Peng Huang$^2$\footnote{huangfp@ihep.ac.cn},
        Yong-Qiang Wang$^1$\footnote{yqwang@lzu.edu.cn},
        Xin-He Meng$^{3,4}$\footnote{xhm@nankai.edu.cn},
        and Yu-Xiao Liu$^1$\footnote{liuyx@lzu.edu.cn, corresponding author}}

\affiliation{
$^1$Institute of Theoretical Physics, Lanzhou University, Lanzhou 730000, China\\
$^2$Theoretical Physics Division, Institute of High Energy Physics,
Chinese Academy of Sciences, P.O.Box 918-4, Beijing 100049, China\\
$^3$School of Physics, Nankai University, Tianjin 300071, China\\
$^4$State Key Laboratory of Theoretical Physics, Institute of Theoretical Physics,
Chinese Academy of Science, Beijing 100190, China}

\begin{abstract}
The future gravitational wave (GW) observations of compact binaries and their possible electromagnetic counterparts may be used to probe the nature of the extra dimension. It is widely accepted that gravitons and photons are the only two completely confirmed objects that can travel along null geodesics in our four-dimensional space-time. However, if there exist extra dimensions and only GWs can propagate freely in the bulk, the causal propagations of GWs and electromagnetic waves (EMWs) are in general different. In this paper, we study null geodesics of GWs and EMWs in a five-dimensional anti-de Sitter space-time in the presence of the curvature of the universe. We show that for general cases the horizon radius of GW is longer than EMW within equal time. Taking the GW150914 event detected by the Advanced Laser Interferometer Gravitational-Wave Observatory and the X-ray event detected by the Fermi Gamma-ray Burst Monitor as an example, we study how the curvature $k$ and the constant curvature radius $l$ affect the horizon radii of GW and EMW in the de Sitter and Einstein-de Sitter models of the universe. This provides an alternative method for probing extra dimension through future GW observations of compact binaries and their electromagnetic counterparts.
\end{abstract}


\maketitle

\section{Introduction}\label{sec0}

Since Einstein predicted that there exist a special kind of wave solutions (named gravitational waves (GWs)) in linearized weak-field equations \cite{Einstein:1916cc,Einstein:1918btx}, people have never stopped detecting GWs. In 1962, M. E. Gertsenshtein and V. I. Pustovoit proposed the first methodology to detect very long wavelength GWs by using interferometers. Six years later, J. Weber declared that GWs had been detected, but it was proved to be false later. Although it was just a small episode in the history of detecting GWs, this was exactly the seed of the Advanced Laser Interferometer Gravitational-Wave Observatory (LIGO). Since then several scientific organizations started to prepare for new GW detectors and the great achievement did not appear until 2015.

It is really an encouraging  news that LIGO recently observed a transient GW signal, which was also named as GW150914~\cite{Abbott2016blz,TheLIGOScientific2016,Abbott2016gcq,Abbott:2016iqz,
TheLIGOScientific:2016htt}. According to the analysis given by the LIGO scientific collaboration and Virgo collaboration~\cite{Accadia:2012zzb}, this GW signal came from the merger of a pair of black holes and the luminosity distance between the source of the GW150914 and our earth is about 430~Mpc (with redshift $z\sim0.09$). Since LIGO announced the first direct detection of GWs, the research fruits relative to it have sprung up. With the GW150914, one can impose restrictions on cosmological viscosity~\cite{Goswami:2016tsu}, hunt for dark particles~\cite{Giudice:2016zpa}, and detect triple systems~\cite{Meiron:2016ipr} (more applications can be seen in Refs.~\cite{Cardoso:2016rao,Hartwig:2016nde,
Sousa:2016ggw,Lasky:2016knh,Yamamoto:2016bxj,Domcke:2016mbx,Maselli:2016ekw}).

Coincidentally, the Fermi Gamma-ray Burst Monitor (GBM) detected an electromagnetic counterpart after LIGO detected the GW signal, which lasted for about 1 s and appeared about 0.4 s after the GW signal~\cite{Connaughton:2016umz}. According to the data from the Fermi GBM satellite, {the gamma-ray burst  could be related to the GW signal.} In Ref.~\cite{Gogberashvili1602.06526}, the size of the spherical brane-universe expanding in multi-dimensional space-time was constrained by using those data. Although the correlation between the GW signal and electromagnetic wave (EMW) signal needs more observation data (especially, more GW signals to be detected) to be confirmed, there are not a few works trying to explain why the arrival time is different between the GW and EMW signals~\cite{Bagoly:2016ytk,Grado:2016sor,Cowperthwaite:2016shk,Ellis:2016rrr,
Abbott:2016iqz,Greiner:2016dsk,Branchina:2016gad,Janiuk:2016qpe,Morokuma:2016hqx,
Ho:2016qqm,Lipunov:2016cml,Lipunov:2016kmy,Tavani:2016jrd}. In Ref.~\cite{Takahashi:2016jom}, the authors drew a conclusion that the GW can arrive earlier than the EMW if the two signals are emitted simultaneously and pass through a lens during the travel. However, there are also some literatures which are not agree with this correlation between these two signals~\cite{Zhang:2016kyq,Xiong:2016ssy}.

{In this work, we take these two events as an example to study whether the arrival time difference 0.4~s is reasonable in our extra dimension model, in which our universe is supposed to be a three-brane embedded in a five-dimensional Schwarzschild-anti-de Sitter space-time (more other applications of GWs with extra dimensions refer to Refs.~\cite{Clarkson:2006pq,Garcia-Aspeitia:2013jea,Garcia-Bellido:2016zmj}). It is worth noting that although these two events are not necessarily related, we just take them as an example to illustrate how to analyze the time difference between GW signals and their counterparts with the variation of parameters. Therefore, whether the X-ray is related to GW150914 is not important in our study and our research is mainly for the future accurate measurements of GW events and corresponding electromagnetic counterparts (for example, the merging of neutron star binaries). Theoretically, the electromagnetic counterparts involving the merging of compact binaries are much more promising to be observed. The generation mechanism of the electromagnetic counterparts and the simultaneity of these two kinds of signals are both interesting research subjects and worth exploring thoroughly in the future~\cite{Branchina:2016gad,Connaughton:2016umz,Vachaspati:2007fc,Li:2016iww,Morsony:2016upv,Loeb:2016fzn,
Fermi-LAT:2016qqr,Racusin:2016fko}.}

In general extra dimension models, gravity can travel in the bulk space, but fermions and gauge boson fields are trapped on the brane. Therefore, gravitons always travel along the shortest null paths in the bulk, which are also called ``shortcuts'' ~\cite{Caldwell:2001ja,Ishihara:2000nf,Abdalla:2001he,Abdalla:2005wr,Abdalla:2002je,
Abdalla:2002ir,Pas:2005rb}. According to Ref.~\cite{Caldwell:2001ja}, if the constant curvature $k$ vanishes, the gravitational horizon radius is always larger than or equal to the horizon radius of EMW (it depends on the constituents of our universe) in a five-dimensional Schwarzschild-anti-de Sitter space-time. But to obtain an observable time delay between EMW and GW, the luminosity distance of their source needs to be incredibly long. We make our study on the foundation of Ref.~\cite{Caldwell:2001ja} and take the constant curvature $k$ into account. Our research shows that although the curvature $k$ is a tiny parameter, it could cause an observable time delay between EMW and GW even if the source of the signals is ``near'' to us. This observable effect provides a new method for probing extra dimension and constraining the parameters of extra dimension. We consider two general models of the universe and find that this five-dimensional Schwarzschild-anti-de Sitter model, indeed, can explain the arrival time difference between the EMW and GW signals in the GW150914 event without violating current observational data.

In section~\ref{sec1}, the null geodesics for gravitons and photons are given in the five-dimensional anti-de Sitter (AdS) space-time. In section~\ref{sec2}, the de Sitter model is considered. In section~\ref{sec3}, the Einstein-de Sitter model is discussed. The final conclusions and discussions are shown in section~\ref{sum}.

\section{Null geodesics in five-dimensional AdS space-time}
\label{sec1}

We intend in this section to review how to obtain the null geodesics through the bulk in five-dimensional AdS space-time and the null geodesics on a brane. The general Schwarzschild-anti-de Sitter metric ansatz we will study is
\begin{eqnarray}\label{Metric1}
ds^2=-f(R)dT^2+f(R)^{-1}dR^2+R^2d\Sigma_{k}^{2},
\end{eqnarray}
where $d\Sigma_{k}^{2}$ is a metric on a locally homogeneous three-dimensional surface of constant curvature $k$ ($k=+1$, $k=0$, and $k=-1$ correspond to three-dimensional sphere, flat, and hyperboloid, respectively)
\cite{Birmingham:1998nr,Mann:1996gj,Brill:1997mf,Vanzo:1997gw}. The coordinate $T$ is referred as Killing time and the coordinates $(T,~R)$ are collectively called curvature coordinates \cite{Brill:1997mf}. The function $f(R)$ takes the form
\begin{eqnarray}\label{fR}
f(R)=k+\frac{R^2}{l^2}-\frac{M}{R^2}.
\end{eqnarray}
Here, $l$ ($>0$) is a parameter with dimension(s) of length, which is also called the constant curvature radius of the five-dimensional AdS space-time. The parameter $M$ is the Schwarzschild-like mass. In Ref.~\cite{Caldwell:2001ja}, the authors considered the case $k=M=0$ and obtained the geodesics in the bulk and on a brane. To investigate how the curvature of the universe influences these null geodesics, we briefly derive these geodesics in the case $k\neq0$ and $M=0$.

Before the process of the derivation, there are two assumptions that should be stated. The first one is that the bulk is supposed to be empty,  although  it  generally contains multiple fields (these fields are negligible because the energy of these fields is insignificant compared to the negative cosmological constant). Besides that, there may be multiple branes in the bulk, but if they are sufficiently distant from each other we can still assume the bulk is empty. The second one is that the three-brane which we concern is homogeneous and isotropic. Actually, if the brane represents our universe, this assumption should be reasonable.

In the sight of observers on any three-brane, the orbit of the corresponding three-brane is given by $R(T)$. The function $R(T)$, which is simply determined by the position of the brane in the extra dimension, is reinterpreted as the scale factor $a(t)$ on the three-brane with proper time $t$. We suppose the orbit of our universe also depends on the function $R(T)$, thus the proper time $t$ for the comoving observers can be defined as
\begin{eqnarray}\label{propert}
dt^2=f(R)dT^2-\frac{dR^2}{f(R)}.
\end{eqnarray}
Combining this definition with Eq.~(\ref{Metric1}) one can obtain the induced metric on our three-brane
\begin{eqnarray}\label{Metric2}
ds^{2}_{brane}=-dt^2+R(t)^2d\Sigma_{k}^{2}.
\end{eqnarray}
Note that the usual cosmological scale factor $a(t)$ on our brane is equivalent to $R(t)$, i.e., $R(t)\equiv a(t)$.

\begin{figure}[htb]
\begin{center}
\includegraphics[width=12cm]{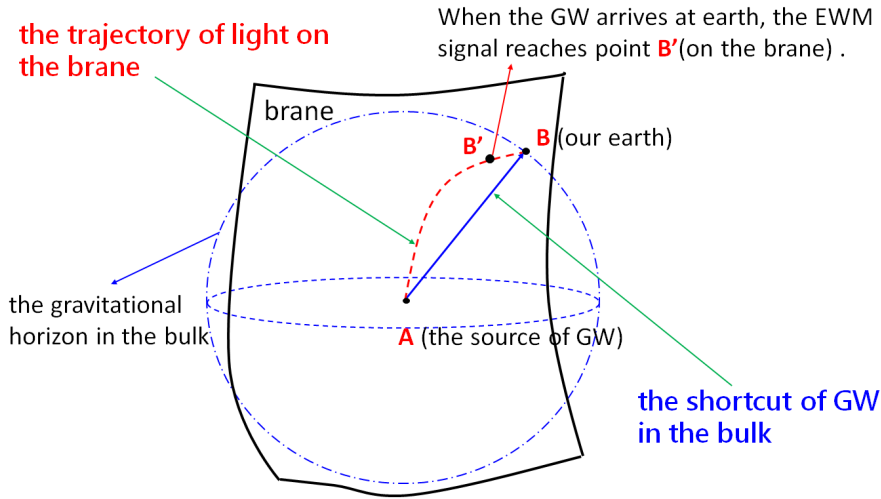}
\end{center}
\caption{\label{shiyitu} The geodesics of the EMW and GW. The points A, B, and B$'$ are all on the brane. The dashed red line AB$'$B represents the track of the null geodesic on the brane and the solid blue line AB is the track of the null geodesic in the bulk.}
\end{figure}

Now, we simply illustrate the physical picture involved in the following investigation. Referring to the methodologies in Refs.~\cite{Caldwell:2001ja,Mukohyama:1999wi}, one can introduce a spherical coordinate system $(r, \theta, \phi)$ on the brane and consider one point A on the brane as the origin of coordinates (see Fig.~\ref{shiyitu}). We assume there is a GW signal emitted by the initial point A and received by the point B which is also on the brane. Therefore, the track of the GW is a radial null geodesic though the bulk. We can ignore the angular variables $(\theta, \phi)$ and set a three-dimensional metric to describe the propagation of the GW in the bulk
\begin{eqnarray}\label{Metric3}
ds^{2}=-f(R)dT^2+f(R)^{-1}dR^2+R^2dr^2.
\end{eqnarray}
For this metric, since the coordinates $T$ and $R$ are independent of the radial coordinate $r$, the vector $(\frac{\partial}{\partial r})^a$ is apparently a Killing vector. In addition, the vector $(\frac{\partial}{\partial T})^a$ is also a Killing vector for $f(R)>0$. To obtain the geodesic trajectory, one can define a unit vector $k^a=\frac{d x^a}{d \lambda}$ ($g_{\mu\nu}k^{\mu}k^{\nu}=1$), which is always tangent to the geodesic. Based on this, combining with the Killing vectors $(\frac{\partial}{\partial T})^a$ and $(\frac{\partial}{\partial r})^a$, the tangent vector $k^a$ should satisfy
\begin{eqnarray}\label{killF1}
g_{\mu\nu}k^{\mu}(\frac{\partial}{\partial T})^{\nu}=-f(R)\frac{dT}{ d\lambda}\equiv k_T,
\end{eqnarray}
and
\begin{eqnarray}\label{killF2}
g_{\mu\nu}k^{\mu}(\frac{\partial}{\partial r})^{\nu}=R^2\frac{dr}{d\lambda}\equiv{k_r},
\end{eqnarray}
where $k_T$ and $k_r$ are two constants along the geodesic. Besides, the null geodesic leads to $ds^2=0$. From Eq.~(\ref{Metric3}), we have
\begin{eqnarray}\label{ds=0}
0=-f(R)^2dT^2+dR^2+f(R)R^2dr^2.
\end{eqnarray}
Dividing the entire expression~(\ref{ds=0}) by $d\lambda ^2$ and using Eqs.~(\ref{killF1}) and (\ref{killF2}) to eliminate $\frac{dT}{ d\lambda}$ and $\frac{dr}{d\lambda}$, we get
\begin{eqnarray}
\left(\frac{dR}{d\lambda}\right)^2=k_T^2-k_r^2\frac{f(R)}{R^2}.
\end{eqnarray}
Note that the parameter $d\lambda$ still exists. One can employ $k_r$ and $k_T$ again, respectively, to eliminate the parameter $d\lambda$ in the above equation. Therefore, it is easy to get the trajectories of the infinitesimal version geodesic along the radial $r$ and time $T$, which are given as
\begin{eqnarray}\label{dr}
\left(\frac{k_T^2}{k_r^2}-\frac{f}{R^2}\right)^{-1/2}\frac{dR}{R^2}=dr
\end{eqnarray}
and
\begin{eqnarray}\label{dt}
\frac{dR}{f\sqrt{1-\frac{k_r^2 f}{k_T^2 R^2}}}=dT.
\end{eqnarray}
Substituting $f(R)=k+\frac{R^2}{l^2}$ into Eqs.~(\ref{dr}) and (\ref{dt}), respectively, and integrating them from the initial point A to the point B, we get
\begin{eqnarray}\label{rba}
r_{AB}\equiv\int_{r(A)}^{r(B)} dr
&=&\int_{R(A)}^{R(B)} \left(\frac{k_T^2}{k_r^2}-\frac{f}{R^2}\right)^{-1/2}\frac{1}{ R^2}dR\nonumber\\
&=&\frac{1}{\sqrt{k}}\Bigg[\arctan\left(\sqrt{\frac{k}{Q R(A)^2-k}}\right)
-\arctan\left(\sqrt{\frac{k}{Q R(B)^2-k}}\right)\Bigg],
\end{eqnarray}
and
\begin{eqnarray}\label{tba}
T_{AB}\equiv\int_{T(A)}^{T(B)}dT
&=&\int_{R(A)}^{R(B)}\frac{dR}{f\sqrt{1-\frac{k_r^2 f}{k_T^2 R^2}}}\nonumber\\
&=&\frac{l}{\sqrt{k}}\Bigg[\arctan\left(\sqrt{\frac{QR(B)^2-k}{k{l^2}/s}}\right)
-\arctan\left(\sqrt{\frac{QR(A)^2-k}{k{l^2}/s}}\right)\Bigg],
\end{eqnarray}
where $Q=\frac{1}{s}-\frac{1}{l^2}$ and $s=(\frac{k_r}{k_T})^2$ are constants. Note that $R(A)$ and $R(B)$ are the values of the scale factor $a(t)$ at the space-time points A and B for the observers on the brane, respectively. Since we only concern about observations on the brane, we identify $a(t_A)$ and $a(t_B)$ with $R(A)$ and $R(B)$. Although we have expressed $r_{AB}$ in terms of the parameters on the brane, we do not know how to determine the value of the parameter $s=(\frac{k_r}{k_T})^2$ clearly. Therefore, making use of Eqs.~(\ref{rba}) and (\ref{tba}) to eliminate $s$, we have
\begin{eqnarray}\label{rgab}
r_{AB}
=\frac{1}{\sqrt{k}}\arctan[\frac{\sqrt{k}}{\sqrt{-k+\frac{a(t_A)^2\big[kG_a
G_b(G_a+G_b)\csc(\overline{T}_{AB})^2
-2\csc(\overline{T}_{AB})^2\sqrt{k^2G_a^3G_b^3\cos (\overline{T}_{AB})^2}\big]}
{G_a^2G_b^2-l^2\big[kG_aG_b(G_a+G_b)\csc(\overline{T}_{AB})^2
-2\csc(\overline{T}_{AB})^2\sqrt{k^2G_a^3G_b^3\cos (\overline{T}_{AB})^2}\big]}}}]\nonumber\\
-\frac{1}{\sqrt{k}}\arctan[\frac{\sqrt{k}}{\sqrt{-k+\frac{a(t_B)^2\big[kG_a
G_b(G_a+G_b)\csc(\overline{T}_{AB})^2
-2\csc(\overline{T}_{AB})^2\sqrt{k^2G_a^3G_b^3\cos (\overline{T}_{AB})^2}\big]}
{G_a^2G_b^2-l^2\big[kG_aG_b(G_a+G_b)\csc(\overline{T}_{AB})^2
-2\csc(\overline{T}_{AB})^2\sqrt{k^2G_a^3G_b^3\cos (\overline{T}_{AB})^2}\big]}}}],
\end{eqnarray}
where $\overline{T}_{AB}=\frac{\sqrt{k}T_{AB}}{l}$, $G_a=kl^2+a(t_A)^2$, and $G_b=kl^2+a(t_B)^2$.

Now, getting back to the definition of the proper time $t$ in Eq.~(\ref{propert}) and the expression of $f(R)$, we get the relation between the parameter $T$ and the proper time $t$ on the brane
\begin{eqnarray}\label{dt2}
dT=\frac{\sqrt{k+\frac{a(t)^2}{l^2}+\dot{a}(t)^2}}{k+\frac{a(t)^2}{l^2}}dt.
\end{eqnarray}
With any given specific expression of the scale factor $a(t)$, $T_{AB}$ can be calculated through Eq.~(\ref{dt2}), which is not the function of the parameter $s$. Putting the result of $T_{AB}$ into Eq.~(\ref{rgab}), we obtain the final expression of $r_{AB}$, which is also called gravitational horizon radius in Ref.~\cite{Caldwell:2001ja}. Note that we get $r_{AB}$ in terms of the quantities on the brane, but the null geodesic traveling from the point A to B is in the bulk.

As for the null geodesic on the brane, it is always characterized by EMW signal. The corresponding horizon radius is given by
\begin{eqnarray}
\int_{r_A}^{r_{B'}}\frac{1}{\sqrt{1-kr^2}}dr=\int_{t_A}^{t_B}\frac{1}{a(t)}dt.
\end{eqnarray}
We define $\widetilde{r}_{AB'}$ as the horizon radius of the EMW and get the following abstract expression of $\widetilde{r}_{AB'}$,
\begin{eqnarray}
\widetilde{r}_{AB'}=\frac{1}{\sqrt{k}}
\sin\left[\sqrt{k}\int_{t_A}^{t_B}\frac{1}{a(t)}dt\right].
\end{eqnarray}

Here, we suppose that the EMW and GW are emitted simultaneously at the time $t_A$ from the same spatial point A (on the brane). The GW signal traverses in the bulk and is received at the spatial point B (on the brane). The radial distance of this trajectory is ${r}_{AB}$  between times $t_A$ and $t_B$  . The EMW signal gets to the spatial point B$'$ (on the brane) and travels a radial distance $\widetilde{r}_{AB'}$ between times $t_A$ and $t_B$ (see Fig.~\ref{shiyitu}).

\section{De Sitter model}
\label{sec2}
In this section, we give a concrete comparison of $r_{AB}$ and $\widetilde{r}_{AB'}$ by  considering the de Sitter model of the universe, in which the universe is dominated by constant vacuum energy. The fundamental Friedmann equation is given by
\begin{eqnarray}\label{friedmann1}
\dot{a}(t)^2+k={N_{\Lambda}} a(t)^2,
\end{eqnarray}
where $N_{\Lambda}=\frac{8\pi\rho_\Lambda G}{3}$ is a constant and the dot denotes the derivative with respect to $t$. According to the Friedmann equation we have $dt=\frac{1}{\sqrt{N_{\Lambda}a(t)^2-k}}da(t)$. Plugging it into Eq.~(\ref{dt2}), performing the integration from the point A ($a(t_A)$) to B ($a(t_B)$) and using the Friedmann equation (\ref{friedmann1}) to eliminate the parameter $N_{\Lambda}$, we obtain
\begin{eqnarray}\label{TAB}
T_{AB}=\frac{1}{\sqrt{k}}\left[\arctan\Bigg(\frac{H_B}{\sqrt{\widetilde{k}+
\widetilde{k}l^2(H_B^2+\widetilde{k})}}\Bigg)
-\arctan\Bigg(\frac{\sqrt{H_B^2+\widetilde{k}-(1+z)^2\widetilde{k}}}
{(1+z)\sqrt{\widetilde{k}+\widetilde{k}l^2(H_B^2+\widetilde{k})}}\Bigg)\right].
\end{eqnarray}
Here, $\widetilde{k}=\frac{k}{a(t_B)^2}$, $H_B=\frac{\dot{a}(t_B)}{a(t_B)}$ is the value of the Hubble parameter at time $t_B$, and $1+z=\frac{a(t_B)}{a(t_A)}$.

Analogously, we can utilize the Friedmann equation to get the specific $r_{AB}$ and $\widetilde{r}_{AB'}$. In order to simplify them, we only consider the case of $k\geq0$ (this guarantees $G_a>0$ and $G_b>0$ in Eq.~(\ref{rgab})). Then $r_{AB}$ and $\widetilde{r}_{AB'}$ are given by
\begin{eqnarray}\label{rgAB}
r_{AB}=\frac{1}{ \sqrt{k}}
\arctan\Bigg[\frac{1}{\sqrt{-1+\frac{(1+z)^{-2}\big[(\widetilde{G}_a+\widetilde{G}_b)
\csc(\overline{T}{_{AB}})^2
-2\csc(\overline{T}{_{AB}})^2\sqrt{\widetilde{G}_a\widetilde{G}_b
\cos(\overline{T}{_{AB}})^2}\big]}{\widetilde{G}_a\widetilde{G}_b-
\widetilde{k}l^2\big[(\widetilde{G}_a+\widetilde{G}_b)
\csc(\overline{T}{_{AB}})^2
-2\csc(\overline{T}{_{AB}})^2\sqrt{\widetilde{G}_a\widetilde{G}_b
\cos(\overline{T}{_{AB}})^2}\big]}}}\Bigg]\nonumber\\
-\frac{1}{ \sqrt{k}}
\arctan\Bigg[\frac{1}{\sqrt{-1+\frac{\big[(\widetilde{G}_a+\widetilde{G}_b)
\csc(\overline{T}{_{AB}})^2
-2\csc(\overline{T}{_{AB}})^2\sqrt{\widetilde{G}_a\widetilde{G}_b
\cos(\overline{T}{_{AB}})^2}\big]}{\widetilde{G}_a\widetilde{G}_b-
\widetilde{k}l^2\big[(\widetilde{G}_a+\widetilde{G}_b)
\csc(\overline{T}{_{AB}})^2
-2\csc(\overline{T}{_{AB}})^2\sqrt{\widetilde{G}_a\widetilde{G}_b
\cos(\overline{T}{_{AB}})^2}\big]}}}\Bigg]
\end{eqnarray}
and
\begin{eqnarray}\label{rlAB}
\widetilde{r}_{AB'}
=\frac{1}{\sqrt{k}}\sin\Bigg[\arctan\Bigg(\frac{\sqrt{\widetilde{k}}}
{\sqrt{-\widetilde{k}+(1+z)^{-2}(H_B^2+\widetilde{k})}}\Bigg)
-\arctan\Bigg(\frac{\sqrt{\widetilde{k}}}{H_B}\Bigg)\Bigg],
\end{eqnarray}
where $\widetilde{G}_a=\widetilde{k}l^2+(1+z)^{-2}>0$ and $\widetilde{G}_b=\widetilde{k}l^2+1>0$. It is difficult to compare $r_{AB}$ and $\widetilde{r}_{AB'}$ because there are three uncertain parameters $\widetilde{k}$, $z$, and $H_B$. Fortunately, the gravitational wave event GW150914 detected by LIGO can help us to compare them in certain range of the parameters and examine the validity and reliability of this extra dimension model. According to Ref.~\cite{Abbott2016blz}, the source of the GW signal lies at a luminosity distance of $410^{+160}_{-180}$ Mpc, which corresponds to a redshift $z=0.09^{+0.03}_{-0.04}$.

Note that here the redshift is estimated in the context of standard general relativity (GR). However, in general, when one considers modified gravity or some other effects, the luminosity-redshift relation  will change. For instance, in Ref.~\cite{Kahya:2016prx}, the authors considered the gravitational potential of the mass distribution and their conclusion is that GW150914 will experience a Shapiro delay (about 1800 days) along the line of sight during its $410^{+160}_{-180}$ Mpc (about $1.4\pm0.6$ billion light years) journey.

{According to the original literatures of LIGO scientific and Virgo collaborations~\cite{Abbott2016blz,TheLIGOScientific:2016wfe}, we know that GW observations are insensitive to the redshift~\cite{Schutz:1986gp}. Therefore, the measurements (including redshift and redshifted masses) are mainly dependent on the luminosity distance to the source. To obtain the redshift of GW signal, LIGO scientific and Virgo collaborations need to directly measure the luminosity distance from the GW signal alone. Then they assume the universe is a flat $\Lambda$CDM model with given Hubble parameter and matter density parameter. Through the Friedmann equation, they can derive the corresponding redshift. Since the amplitude of GW is inversely proportional to the comoving distance in the context of standard GR, as long as the amplitude of GW is obtained one can calculate the luminosity distance~\cite{Abbott2016blz,TheLIGOScientific:2016wfe}. In our braneworld model, because of the presence of an additional dimension, the propagation of gravity is different from that of standard GR. Since gravity could spread throughout the five-dimensional space-time, when the scale we consider is less than the size of the extra dimension, from the Gauss's law in $(4+n)$ dimensions the gravitational potential is given by~\cite{ArkaniHamed:1998rs}
\begin{eqnarray}
V(r)\sim\frac{m_1m_2}{M^{n+2}_{Pl(4+n)}}\frac{1}{r^{n+1}},
\end{eqnarray}
where $r$ is the distance that we consider on the brane.
When the distance is much larger than the size of the extra dimension, the gravitational potential will degenerate into the four-dimensional case~\cite{ArkaniHamed:1998rs,Randall:1999ee}. Therefore, in high-dimensional braneworld models, the hypothesis of small extra dimensions is a common way to restore four-dimensional Newtonian gravitational potential. According to current experiments within the solar system~\cite{Long:2002wn,Iorio:2011ab,Linares:2013opa,Tan:2016vwu}, Newtonian gravity is applicable in the sub-millimeter range, which means that the radius of the extra dimension should be less than 0.1 mm. In our braneworld model, this requires that the AdS radius $l$, which stands for the scale of the extra dimension, is small ($l<0.1$ mm).
Thus, for large distance $r\gg l$, the modification from the extra dimension to Newtonian gravitational potential can be neglected.
The amplitude of GW is still inversely proportional to the comoving distance in our braneworld model and the luminosity distance calculated by our model is not much different from the one obtained by standard GR.

In our braneworld model, the modification of the Friedmann equation is also negligible.} The most general form of the four-dimensional Friedmann equation in braneworld models can be written as~\cite{Binetruy:1999hy,Gergely:2003pn,Dabrowski:2002tp,Nagy:2006gs,Keresztes:2006tw,Szabo:2007mj,BouhmadiLopez:2008nf}
\begin{eqnarray}\label{friedmannEQ}
H^2+\frac{k}{a^2}=\frac{\kappa_{(4)^2}}{3}\rho+\frac{\kappa_{(4)^2}}{6\lambda^2}\rho^2
 +\texttt{other~correction~terms},
\end{eqnarray}
where $\rho$ denotes the density of matters on the brane, $\kappa_{(4)}^2=8\pi G_{(4)}$, and $\lambda$ is the brane tension. The other correction terms depend on the specific models and in order to close to the real universe, usually, they are always negligible in the late-time cosmology compared with the first term. As we known, early in cosmic history, the second term in Eq.~(\ref{friedmannEQ}) is truly important, which could have a great influence on the evolution of the scale factor. But when we consider the late-time cosmology, since the density of matters on the brane becomes smaller and smaller, the second term in the right-hand side of Eq.~(\ref{friedmannEQ}) can also be ignored. In other words, in the late-time cosmology, the four-dimensional behavior of the five-dimensional braneworld gravity can degenerate into standard GR. Therefore, although in the context of braneworld gravity the redshift is different from the one decided by GR, comparing with the observation error this gap is negligibly small. {Given the above, in this example, for convenience we could suppose the redshift calculated by our model is unchanged comparing to the one of standard GR. As for other physical quantities, if they can be given based on the redshift, they also have no big changes. For example in the original literatures of LIGO scientific and Virgo collaborations~\cite{Abbott2016blz,TheLIGOScientific:2016wfe}, since there is no intrinsic mass or length scale in vacuum GR and the redshift of observed frequency of the signal can not be distinguished from the rescaling of the masses~\cite{Cutler:1994ys,Echeverria:1989hg,Krolak:1987ofj}, they could measure redshifted masses $m$ by $m=(1+z)m^{source}$.}

In addition, the Fermi GBM recorded an electromagnetic counterpart with 0.4~s delay after LIGO detected the GW signal. Now the general viewpoint about these two signals is that the EMW signal is massively more likely to be associated with gravitational wave event GW150914. Therefore, based on these two events, we can establish a specific physical picture to compare $r_{AB}$ and $\widetilde{r}_{AB'}$. We hypothesize that the source of the two signals is the initial space-time point A, the present earth  is the point B, and the two signals were generated simultaneously. The current observations indicate that $H_B=H_0\sim2.189\times10^{-18}\rm s^{-1}$ and $\widetilde{k}=\frac{\Omega_kH_0^2}{c^2}\leq10^{-40}\rm s^{-2}$ ($c=1$).

\begin{figure}[htb]
\begin{center}
\includegraphics[width=11cm,height=7cm]{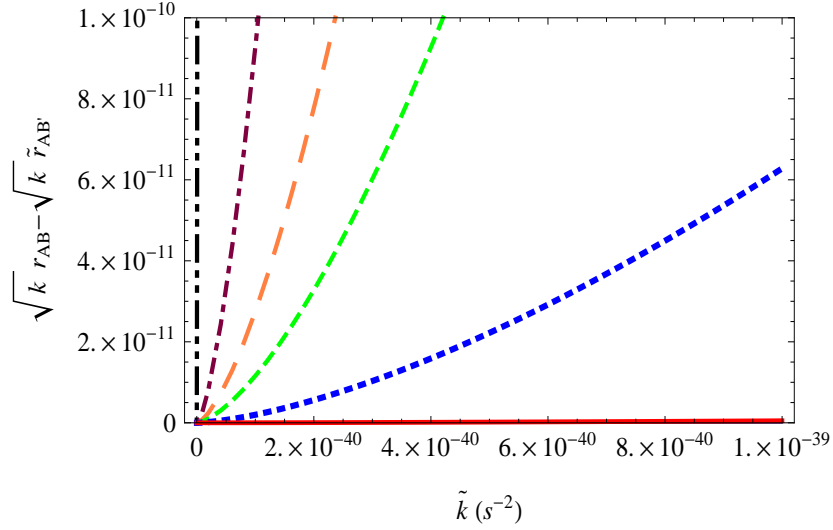}
\caption{\label{chazhitu} The plot of the differences between the horizon radii of the GW and EMW in the de Sitter model. The five sets of data, from top to bottom, are $z=(2, 0.18, 0.12, 0.09, 0.05, 0.01)$, respectively.}
\end{center}
\end{figure}

In this model, Eq.~(\ref{rgAB}) can be further reduced to
\begin{eqnarray}\label{rgAB1}
r_{AB}=\frac{1}{ \sqrt{k}}\Bigg[\arctan\Bigg(\frac{\sqrt{\widetilde{k}}(1+z)}
{\sqrt{H_B^2-\widetilde{k}z(2+z)}}\Bigg)
-\arctan\Bigg(\frac{\sqrt{\widetilde{k}}}{H_B}\Bigg)\Bigg],
\end{eqnarray}
which is irrelevant to the parameter $l$. When the curvature $k$ approaches to zero, it is easy to verify that $r_{AB}=\widetilde{r}_{AB'}=\frac{z}{H_B}$, which is consistent with the discussion in Ref.~\cite{Caldwell:2001ja}, i.e, if $H_B\geq0$ is a constant and the curvature $k=0$, the horizon radius of the EMW on the brane and the gravitational horizon radius in the bulk would be totally the same~\cite{Ishihara:2000nf}. When $k\neq0$, our calculations show that $r_{AB}$, for general case, is larger than $\widetilde{r}_{AB'}$. Note that here we compare $r_{AB}$ and $\widetilde{r}_{AB'}$ by giving the same interval of time (from $t_A$ to $t_B$). As the values of $z$ and $\widetilde{k}$ increase, the gap between $r_{AB}$ and $\widetilde{r}_{AB'}$ gets bigger (see Fig.~\ref{chazhitu}). The positive differences mean the GW travels farther than the EMW in the same time (the different values of the parameter $z$ correspond to different intervals of time). This result agrees with our physical intuition: the GW (in the bulk) travels ``faster" than or at least as fast as the EMW on the brane and with the increase of the curvature of the brane, the shortcut effect of the GW becomes more manifest.

Let us go back to the GW150914 event. Ignoring the impact of other factors on the spread of the GW and EMW signals, one can give a rough estimate of the curvature $\widetilde{k}$ and the cosmological curvature parameter $\Omega_k$. The method of building mathematical model is easy. According to the previous description, in the GW150914 and X-ray events, we have three important space-time points: the initial point A (the source of the GW and EMW signals), point B (the earth,  the moment when the GW was detected), point C (the earth, the moment when the EMW signal was detected). Therefore, we have $r_{AB}=\widetilde{r}_{AC}=\widetilde{r}_{AB'}+\widetilde{r}_{B'C}$. Referring to Eq.~(\ref{rlAB}), $\widetilde{r}_{AC}$ can be written as
\begin{eqnarray}\label{rlAC}
\widetilde{r}_{AC}
&=&\frac{1}{\sqrt{k}}\sin\Bigg[\arctan\Bigg(\frac{\sqrt{\widetilde{k}}}
{\sqrt{(1+z)^{-2}(H_B^2+\widetilde{k})-\widetilde{k}}}\Bigg)
-\arctan\Bigg(\frac{\sqrt{\widetilde{k}}}{H_B}\Bigg)\nonumber\\
&+&\arctan\Bigg(\frac{\sqrt{\widetilde{k}}}{H_B}\Bigg)
-\arctan\Bigg(\frac{\sqrt{\widetilde{k}}}
{\sqrt{(H_B^2+\widetilde{k})\frac{a(t_C)^2}{a(t_B)^2}-\widetilde{k}}}\Bigg)\Bigg].
\end{eqnarray}
Because the time delay between the points B and C (we define it as $\Delta t$), in general, is relatively small (for example, in the GW150914 and X-ray events the time lag is just about 0.4 s), $\frac{a(t_C)^2}{a(t_B)^2}$ is approximately equal to $(1+\Delta t H_B)^2$. Given this, the sum of the last two terms in Eq.~(\ref{rlAC}) can be regarded as a small variation compared with the sum of the first two. The function $\widetilde{r}_{AC}$ can be expanded in Taylor's series:
\begin{eqnarray}\label{rl1AC}
\widetilde{r}_{AC}
&=&\frac{1}{\sqrt{k}}\sin\Bigg\{\arctan\Big[\frac{\sqrt{\widetilde{k}}}
{\sqrt{(1+z)^{-2}(H_B^2+\widetilde{k})-\widetilde{k}}}\Big]
-\arctan\Big[\frac{\sqrt{\widetilde{k}}}{H_B}\Big]\Bigg\}\nonumber\\
&+&\frac{1}{\sqrt{k}}\Bigg\{\arctan\Big[\frac{\sqrt{\widetilde{k}}}{H_B}\Big]
-\arctan[\frac{\sqrt{\widetilde{k}}}
{\sqrt{(H_B^2+\widetilde{k})(1+\Delta t H_B)^2-\widetilde{k}}}]\Bigg\}\nonumber\\
& &\cos\Bigg\{\arctan\Big[\frac{\sqrt{\widetilde{k}}}
{\sqrt{(1+z)^{-2}(H_B^2+\widetilde{k})-\widetilde{k}}}\Big]
-\arctan\Big[\frac{\sqrt{\widetilde{k}}}{H_B}\Big]\Bigg\}\nonumber\\
&=&\widetilde{r}_{AB'}+\frac{1}{\sqrt{k}}\Bigg\{\arctan\Big[\frac{\sqrt{\widetilde{k}}}{H_B}\Big]
-\arctan[\frac{\sqrt{\widetilde{k}}}
{\sqrt{(H_B^2+\widetilde{k})(1+\Delta t H_B)^2-\widetilde{k}}}]\Bigg\}\nonumber\\
& &\cos\Bigg\{\arctan\Big[\frac{\sqrt{\widetilde{k}}}
{\sqrt{(1+z)^{-2}(H_B^2+\widetilde{k})-\widetilde{k}}}\Big]
-\arctan\Big[\frac{\sqrt{\widetilde{k}}}{H_B}\Big]\Bigg\}.
\end{eqnarray}
Previously, we mentioned that $r_{AB}=\widetilde{r}_{AC}=\widetilde{r}_{AB'}+\widetilde{r}_{B'C}$. Combining with Eq.~(\ref{rl1AC}), we can establish an equation to solve the curvature constant $\widetilde{k}$:
\begin{eqnarray}\label{eq1}
r_{AB}-\widetilde{r}_{AB'}
&=&\frac{1}{\sqrt{k}}\Bigg\{\arctan\Big[\frac{\sqrt{\widetilde{k}}}{H_B}\Big]
-\arctan[\frac{\sqrt{\widetilde{k}}}
{\sqrt{(H_B^2+\widetilde{k})(1+\Delta t H_B)^2-\widetilde{k}}}]\Bigg\}\nonumber\\
& &\cos\Bigg\{\arctan\Big[\frac{\sqrt{\widetilde{k}}}
{\sqrt{(1+z)^{-2}(H_B^2+\widetilde{k})-\widetilde{k}}}\Big]
-\arctan\Big[\frac{\sqrt{\widetilde{k}}}{H_B}\Big]\Bigg\}.
\end{eqnarray}

With any given parameter $z$ and the time lag $\Delta t$, we can obtain the corresponding $\widetilde{k}$  by plugging $H_B=H_0\sim2.189\times10^{-18}\rm s^{-1}$ into   Eq.~(\ref{eq1}). For $\Delta t=0.4$ s, some numerical solutions of $\widetilde{k}$ are listed in  table~\ref{table1}. From table~\ref{table1}, we see that when the range of values for redshift $z$ is 0.01 to 2, the bound of the curvature density $\Omega_k$ is more rigorous than the observations given in Ref. \cite{Ade:2015xua}, $\Omega_k<0.005$. As the redshift $z$ grows, the curvature density $\Omega_k$ gets closer to 0. The table~\ref{table1b} shows the numerical solutions of $\widetilde{k}$ with the change of $\Delta t$ for a given value of redshift $z=0.09$. It is not hard to see that for a given redshift $z$, the bigger the time lag between the GW and EMW is, the bigger the curvature parameter $\Omega_k$ will be.

These results mean that the de Sitter model of the universe combined with higher dimensional space-time can explain why the GW preceded the EMW signal by 0.4~s in the GW150914 and X-ray events (the premise is that they were generated at the same time) and it is not in conflict with the current observational data. And if it is the case, the detections of GWs and their electromagnetic counterparts can even provide more restrictions on the curvature density $\Omega_k$.

\begin{table}
\begin{center}
\begin{tabular}{|c| c| c|}
\hline
~~$z$~~ &      $\widetilde{k}~(s^{-2})$ &    $\Omega_k$ \\
\hline
$0.01$ &      $2.5\times10^{-47}$ &    $\sim -10^{-11}$ \\
\hline
$0.05$ &      $2.0\times10^{-49}$  &    $\sim -10^{-13}$\\
\hline
$0.09$ &      $3.5\times10^{-50}$ &    $\sim -10^{-14}$\\
\hline
$0.12$ &      $1.5\times10^{-50}$ &    $\sim -10^{-14}$\\
\hline
$0.18$ &      $4.3\times10^{-51}$ &    $\sim -10^{-15}$\\
\hline
$2$ &      $3.1\times10^{-54}$  &   $\sim -10^{-18}$\\
\hline
\end{tabular}
\end{center}
\caption{ Some numerical solutions of $\widetilde{k}$ and the corresponding cosmological curvature parameter $\Omega_k$ with given $\Delta t=0.4$ s. According to the table above, we see that $\Omega_k$ becomes smaller as the redshift $z$ grows.}\label{table1}
\end{table}

\begin{table}
\begin{center}
\begin{tabular}{|c| c| c|}
\hline
~~$\Delta t~(s)$~~ &      $\widetilde{k}~(s^{-2})$ &    $\Omega_k$ \\
\hline
$0.01$ &      $8.6\times10^{-52}$  &    $\sim -10^{-16}$\\
\hline
$0.4$ &      $3.5\times10^{-50}$ &    $\sim -10^{-14}$ \\
\hline
$10$ &      $8.6\times10^{-49}$ &    $\sim -10^{-13}$\\
\hline
$500$ &      $4.3\times10^{-47}$ &    $\sim -10^{-11}$\\
\hline
\end{tabular}
\end{center}
\caption{ Some numerical solutions of $\widetilde{k}$ and the corresponding cosmological curvature parameter $\Omega_k$ with given $z=0.09$. According to the table above, we see that $\Omega_k$ becomes larger as the time lag $\Delta t$  grows.}\label{table1b}
\end{table}

\section{ Einstein-de Sitter model}
\label{sec3}

In this section, we consider  the Einstein-de Sitter model of the universe, in which the universe is dominated by non-relativistic matter. The Friedmann equation is
\begin{eqnarray}\label{friedmann2}
\dot{a}(t)^2+k={N_m} a(t)^{-1},
\end{eqnarray}
where ${N_m}=\frac{8\pi G \rho_0}{3 a(t_0)^3}$ is a constant. The analyses are similar to the case of vacuum energy. We start with $T_{AB}$, which is given by
\begin{eqnarray}\label{TAB2}
T_{AB}
&=&\frac{1}{\sqrt{k}}
\Bigg\{\frac{1}{8\widetilde{k}^{3/4}\sqrt{H_k}}\Bigg[8l\widetilde{k}^{1/4}(\widetilde{k}-H_k)+
8l\widetilde{k}^{3/4}\sqrt{H_k}\text{arctanh}\bigg[\frac{\sqrt{\widetilde{k}}}{\sqrt{H_k}}\bigg]\nonumber\\
&+&\sqrt{8l}(\widetilde{k}^{2/3}l+H_k)\bigg[\arctan\bigg(1-\frac{\sqrt{2}}{\widetilde{k}^{1/4}\sqrt{l}}\bigg)
-\arctan\bigg(1+\frac{\sqrt{2}}{\widetilde{k}^{1/4}\sqrt{l}}\bigg)\bigg]\nonumber\\
&+&\sqrt{2l}(\widetilde{k}^{2/3}l-H_k)\log\bigg(\frac{\sqrt{\widetilde{k}}l-\sqrt{2}\widetilde{k}^{1/4}\sqrt{l}+1}
{\sqrt{\widetilde{k}}l+\sqrt{2}\widetilde{k}^{1/4}\sqrt{l}+1}\bigg)\Bigg]\nonumber\\
&-&\frac{1}{8\widetilde{k}^{3/4}\sqrt{ZH_k}}
\Bigg[8l\widetilde{k}^{1/4}(\widetilde{k}Z-H_k)+
8l\widetilde{k}^{3/4}\sqrt{ZH_k}\text{arctanh}\bigg[\frac{\sqrt{\widetilde{k}Z}}{\sqrt{H_k}}\bigg]\nonumber\\
&+&\sqrt{8lZ}(\widetilde{k}^{2/3}l+H_k)\bigg[\arctan\bigg(1-\frac{\sqrt{2Z}}{\widetilde{k}^{1/4}\sqrt{l}}\bigg)
-\arctan\bigg(1+\frac{\sqrt{2Z}}{\widetilde{k}^{1/4}\sqrt{l}}\bigg)\bigg]\nonumber\\
&+&\sqrt{2lZ}(\widetilde{k}^{2/3}l-H_k)\log\bigg(\frac{\sqrt{\widetilde{k}}l-\sqrt{2}\widetilde{k}^{1/4}\sqrt{lZ}+Z}
{\sqrt{\widetilde{k}}l+\sqrt{2}\widetilde{k}^{1/4}\sqrt{lZ}+Z}\bigg)\Bigg]\Bigg\},
\end{eqnarray}
where $H_k=\widetilde{k}+H_B^2$ and $Z=(1+z)^{-1}$. Here, to obtain this expression we have made some approximations, which rely on $(\widetilde{k}+H_B^2)l^2\ll1$, i.e., $l\ll10^{18}$~m. According to the current experiments of the gravitational force law, the curvature radius of the five-dimensional AdS space-time $l$ is less than $10^{-3}$~m (and so far less than $ 10^{18}$~m). So, our approximations are reasonable. Note that in this model $r_{AB}$ can be also given by Eq.~(\ref{rgAB}) and $T_{AB}$ is given by Eq.~(\ref{TAB2}). Analogously, $\widetilde{r}_{AB'}$ reads as
\begin{eqnarray}
\widetilde{r}_{AB'}
=\frac{1}{\sqrt{k}}\sin\Bigg\{-2\arctan\Big[\frac{\sqrt{\widetilde{k}}}
{\sqrt{(1+z)(H_B^2+\widetilde{k})-\widetilde{k}}}\Big]
+2\arctan\Big[\frac{\sqrt{\widetilde{k}}}{H_B}\Big]\Bigg\}.
\end{eqnarray}

\begin{figure}[htb]
\begin{center}
\includegraphics[width=11cm,height=7cm]{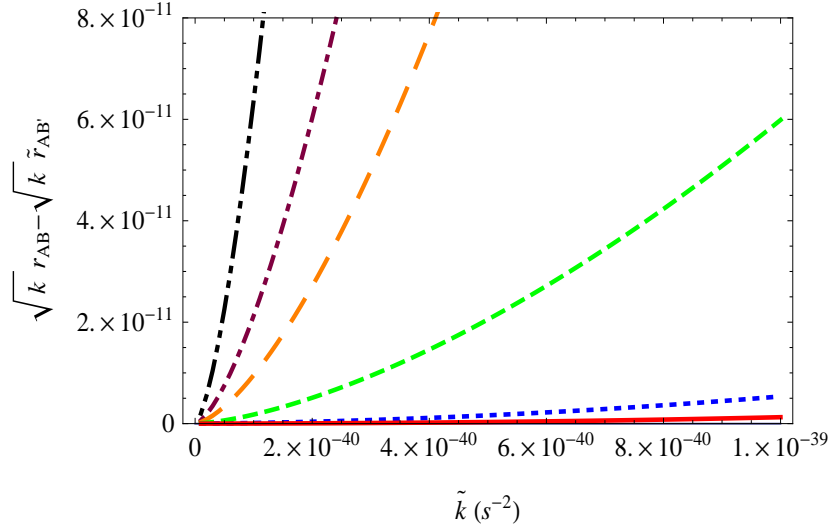}
\caption{\label{chazhitu4} The plot of the differences between the horizon radii of the GW and EMW in the Einstein-de Sitter model. The five sets of data, from top to bottom, are $z=(2, 0.18, 0.12, 0.09, 0.05, 0.01)$. Here, the parameter $l=10^{-5}$~m and we can draw similar conclusions when the parameter $l$ changes.}
\end{center}
\end{figure}

\begin{figure}[htb]
\begin{center}
\includegraphics[width=12cm,height=7cm]{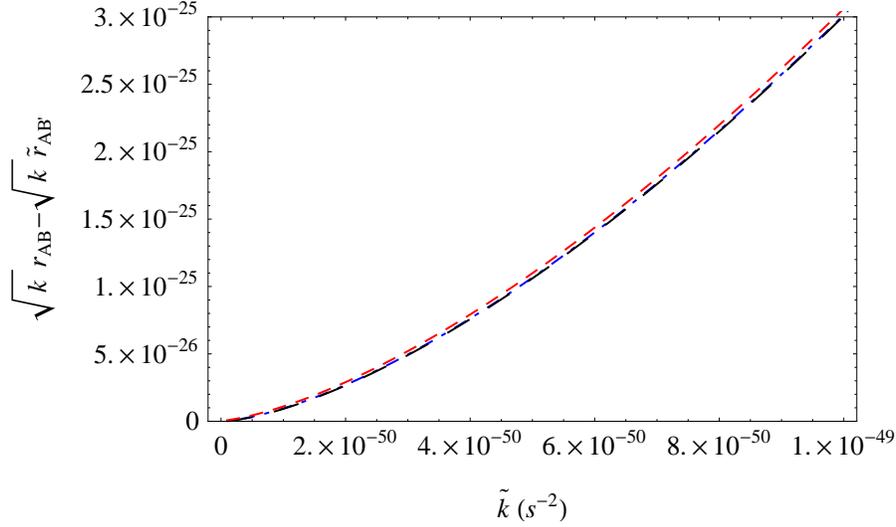}
\caption{\label{chazhitu5} The plot of the differences between the horizon radii of the GW and EMW in the Einstein-de Sitter model. The three values of $l$, from top to bottom, are $l=(10^{-10}~\rm m, 1~\rm m, 10^{10}~\rm m)$. Here, the parameters $z=0.09$ and $\Delta t=0.4$ s. The result indicates that the differences between the horizon radii of the GW and EMW are nearly invariable as the parameter $l$ changes.}
\end{center}
\end{figure}

From Fig.~\ref{chazhitu4}, we see that this solution, indeed, guarantees that $r_{AB}>\widetilde{r}_{AB'}$ for general values of the parameters $z$ and $l$. Finally, we obtain the equation in terms of the parameters $\widetilde{k}$, $l$, $\Delta t$, and $z$, which could be used for solving $\widetilde{k}$. The equation is
\begin{eqnarray}\label{eq2}
r_{AB}-\widetilde{r}_{AB'}
&=&\frac{2}{\sqrt{k}}\Bigg\{-\arctan\Big[\frac{\sqrt{\widetilde{k}}}{H_B}\Big]
+\arctan[\frac{\sqrt{\widetilde{k}}}
{\sqrt{(H_B^2+\widetilde{k})(1+\Delta t H_B)^{-1}-\widetilde{k}}}]\Bigg\}\nonumber\\
& &\times\cos\Bigg\{-2\arctan\Big[\frac{\sqrt{\widetilde{k}}}
{\sqrt{-\widetilde{k}+(1+z)(H_B^2+\widetilde{k})}}\Big]
+2\arctan\Big[\frac{\sqrt{\widetilde{k}}}{H_B}\Big]\Bigg\}.
\end{eqnarray}
Since this equation is more complex than the previous model and involves four arguments, we can choose some representative values of the arguments $l$, $\Delta t$, and $z$ to solve the parameter $\widetilde{k}$ through mapping. For the set of parameters $\Delta t=0.4$ s, $z=0.09$, $l=10^{-10}$~m, $l=1$~m, and $l=10^{10}$~m, we can infer that $r_{AB}$ is not sensitive to the value of $l$ (note that $\widetilde{r}_{AB'}$ is not the function of $l$) (see Fig.~\ref{chazhitu5}). We find that as long as $l$, $\Delta t$, and $z$ take acceptable values (for example $l=10^{-5}$~m, $\Delta t=0.4$~s, and $z$ changes from 0.01 to 2 (see table~\ref{table2})), the corresponding $\widetilde{k}$ is in agreement with the results of observations. With given $z$ and $l$, as the parameter $\Delta t$ changes the trends of the curvature parameter $\Omega_k$ are similar to the de Sitter model (see table~\ref{table2b} as an example).

In Ref.~\cite{Caldwell:2001ja}, the authors discussed this model with the case of $k=0$ in detail. They obtained the approximation ratio of the GW distance to the EMW distance between times $t_A$ and $t_B$, which is given by
\begin{eqnarray}\label{ratio}
\frac{r_g}{r_\gamma}\approx1+\frac{1}{2}(lH_B)^2\frac{1+3\omega}{5+3\omega}
\left(\frac{a_B}{a_A}\right)^{(5+3\omega)/2},
\end{eqnarray}
where $\omega=P/\rho$  is the equation of state of the matter composing the brane. If one considers matter eras ($\omega=0$) and $t_B$ is the present moment, the above equation reduces to
\begin{eqnarray}\label{ratio2}
\frac{r_g}{r_\gamma}\approx1+\frac{1}{10}(lH_0)^2(1+z)^{5/2}.
\end{eqnarray}
Therefore, we can do a rough calculation about the curvature radius $l$ based on the GW150914 event. The calculating result shows that to account for the measured time lag of 0.4~s, the curvature radius $l$ ($\sim10^{18}$~m for $z=0.09$) has to be far greater than the size of the extra dimension, which should be less than $\sim1~\rm mm$ \cite{Hoyle:2000cv}. So, it is observed that only with the consideration of the constant curvature $k\neq0$ in this model (include the de Sitter model) can we explain why the EMW signal arrived earth 0.4~s later than the GW150914 without against the current observational and experimental constraints.

\begin{table}[h!]
\begin{center}
\begin{tabular}{|c| c| c|}
\hline
~~$z$~~ &      $\widetilde{k}~(s^{-2})$ &    $\Omega_k$ \\
\hline
$0.01$ &      $3.1\times10^{-47}$ &    $\sim -10^{-11}$ \\
\hline
$0.05$ &      $2.2\times10^{-49}$  &    $\sim -10^{-13}$\\
\hline
$0.09$ &      $4.2\times10^{-50}$ &    $\sim -10^{-14}$\\
\hline
$0.12$ &      $1.9\times10^{-50}$ &    $\sim -10^{-14}$\\
\hline
$0.18$ &      $6.3\times10^{-51}$ &    $\sim -10^{-15}$\\
\hline
$2$ &         $4.2\times10^{-53}$  &   $\sim -10^{-17}$\\
\hline
\end{tabular}
\end{center}
\caption{Some numerical solutions of $\widetilde{k}$ and the corresponding cosmological curvature parameter $\Omega_k$ with given $\Delta t=0.4$ s and $l=10^{-5}$ m. The result shows that $\Omega_k$ also becomes smaller and smaller in this model as the parameter $z$ grows.}\label{table2}
\end{table}

\begin{table}
\begin{center}
\begin{tabular}{|c| c| c|}
\hline
~~$\Delta t~(s)$~~ &      $\widetilde{k}~(s^{-2})$ &    $\Omega_k$ \\
\hline
$0.01$ &      $1.2\times10^{-51}$  &    $\sim -10^{-15}$\\
\hline
$0.4$ &      $4.2\times10^{-50}$ &    $\sim -10^{-14}$ \\
\hline
$10$ &      $1.2\times10^{-48}$ &    $\sim -10^{-12}$\\
\hline
$500$ &      $5.4\times10^{-47}$ &    $\sim -10^{-11}$\\
\hline
\end{tabular}
\end{center}
\caption{Some numerical solutions of $\widetilde{k}$ and the corresponding cosmological curvature parameter $\Omega_k$ with given $z=0.09$ and $l=10^{-3}$~m.}\label{table2b}
\end{table}

\section{Conclusions and Discussions}
\label{sum}

The exciting discovery of the GW150914 and GW151226~\cite{Abbott:2016nmj} by LIGO opens a new era of multi-messenger astronomy, which will achieve regular detections of GWs from merging compact binaries, and further pursue the electromagnetic counterparts in the multi-wavelength campaigns. The most promising electromagnetic counterpart of compact binaries involves neutron stars. The precise detection of the electromagnetic counterpart will help to understand the properties of the GW sources, such as the redshift of the GW sources, and it can also be used to unravel the effects of extra dimensions.

In this paper, we have taken the GW150914 and X-ray events as an example to show how to constrain the parameters of the models with one extra dimension, although it is not sure whether this gamma ray with 0.4~s  time delay is the true electromagnetic counterpart at present.

In general, we have calculated two kinds of null geodesics for the GW and EMW in the background of a five-dimensional AdS space-time with the case $k\neq0$ and $M=0$. Based on a series of assumptions and simplifications, the general null geodesic of the GW is given by Eq.~(\ref{rgab}), which is the function of the curvature radius $l$, the constant curvature $k$, the Hubble parameter $H_B$, and the cosmological redshift parameter $z$. According to the fact that the GW arrived earlier than the EMW in the GW150914 event, we studied the differences between the horizon radii of the GW and EMW signals due to the effects of extra dimension in the de Sitter model of the universe. We found that the solution (\ref{rgab}) is not the function of the curvature radius $l$. In addition, the horizon radii of the GW and EMW are exactly equal when the constant curvature $k$ approaches to zero. These results are all consistent with the discussion in Ref.~\cite{Caldwell:2001ja}. Compared with the case of $k=0$ in the de Sitter model, the significant distinction of the case for $k\neq0$ is that the horizon radius of the GW is larger than the EMW between times $t_A$ and $t_B$. Combining the GW150914 event with our calculation, we found that if we assume the two signals were generated simultaneously, it is possible to explain, by this extra dimension model, why the GW signal arrived 0.4~s earlier than the EMW signal. In the Einstein-de Sitter model, although the horizon radius of the GW becomes related to the curvature radius $l$, our results show that in this model the GW signal can also be detected earlier than the EMW signal.

As more data with high precision are accumulated by the GW detectors \footnote{As we known, more and more GW detectors are planned, which include eLISA (the Evolved Laser Interferometer Space Antenna), KAGRA (the KAmioka GRAvitational wave detector), DECIGO (the DECi-hertz Interferometer Gravitational wave Observatory), Taiji and TianQin (a proposal for a space-borne detector of GW), etc.} and GBM in the future, more GW signals of compact binaries will be detected. Especially, the GW events of binary neutron stars or neutron star-black hole mergers and the associated electromagnetic counterparts are also expected to be observed. The possible time separation between GWs and their EMW counterparts can provide an alternative way \footnote{The traditional experiments of probing extra dimensions are the precise tests of the gravitational force law and the collider signals of extra dimensions.} to explore the nature of extra dimension, since the gravity can take shortcuts in extra dimensions. Once the exact time separation between the GW signals of compact binary mergers and their electromagnetic counterparts are observed in the future, our study on the differences of the GW and EMW horizon radii in the extra dimension models in this paper can be directly applied to unravel the nature of extra dimension.

Finally, we briefly discuss what the impacts on our results are when one considers a realistic cosmology. {For an inhomogeneous and/or anisotropic four-dimensional cosmology, the propagation of light will be greatly affected and the impact on gravity is not very clear in most cases. The representative discussions in four-dimensional universe can be found in Refs.~\cite{Takahashi:2016jom,Desai:2008vj}. In the context of braneworld gravity with inhomogeneous and/or anisotropic brane, the influence on the bulk geodesics from the inhomogeneity and anisotropy may be important. Especially when scale grows large, the effect may even accumulate. Whether there will be a great effect when the brane is inhomogeneous and/or anisotropic is worthy of our study in the future. If the inhomogeneity and anisotropy of the brane indeed lead to additional delay between the two signals, then the restrictions on the extra dimension models would be more stringent after removing this part of the delay.}

Analogously, for our realistic universe, which consists of dark matter, dark energy, and ordinary matter, the evolution of the scale factor $a(t)$ is more complex. In addition, how each ingredient of the universe affects the spreads of the GW and EMW is not entirely clear~\cite{Desai:2008vj}. According to Refs.~\cite{Takahashi:2016jom,Desai:2008vj}, we expect that the impact of the realistic inhomogeneous universe on the spreads of the GW and EMW for a short distance is larger than that of the background spatial curvature $k$. And for a long distance, the latter might become increasingly important. Of course, the final conclusions need more later studies.

\section*{Acknowledgments} \hspace{5mm}
This work was supported by the National Natural Science Foundation of China (Grants Nos. 11522541, 11375075 and 11205074), and the Fundamental Research Funds for the Central Universities (Grant Nos. lzujbky-2015-jl1 and lzujbky-2016-k04). F.P.H. is supported by the NSFC under grants Nos. 11121092,
11033005, 11375202, the CAS pilotB program and the China Postdoctoral Science Foundation under Grant No. 2016M590133.

\providecommand{\href}[2]{#2}\begingroup\raggedright

\end{document}